\documentclass[letterpaper]{JHEP3}
\pdfoutput=1

\usepackage{latexsym}
\usepackage{amsmath, mathrsfs}
\usepackage{graphicx, slashed}

\title{Pseudo-Dirac Dark Matter Leaves a Trace}

\author{Andrea De Simone\\
        Center for Theoretical Physics, 
        Massachusetts Institute of Technology, Cambridge, Massachusetts - USA\\
        \emph{E-mail:} \email{andreads@mit.edu}}
\author{Veronica Sanz and Hiromitsu Phil Sato\\
        Department of Physics and Astronomy, York University, Toronto, Ontario - Canada\\
        \emph{E-mail:} \email{vsanz@yorku.ca}, \email{satohiro@yorku.ca}}

\abstract{
Pseudo-Dirac Dark Matter is a viable type of dark matter  which originates from a new Dirac fermion whose two Weyl states get slightly split in mass by a small Majorana term.
The decay of the heavier to the lighter state naturally occurs over a detectable length scale.
Thus, whenever pseudo-Dirac Dark Matter is produced in a collider,  it leaves a clear trace: a visible displaced vertex in association with missing energy.
Moreover, pseudo-Dirac Dark Matter  behaves Dirac-like for relic abundance and Majorana-like in direct detection experiments:  it has efficient s-wave annihilations  
but it lacks of dangerous vector interactions with the quarks in the nuclei. 
We provide a general treatment using an effective field theory approach, then specializing 
to the supersymmetric situation of a pseudo-Dirac Bino.
The dark matter mass and the mass splitting
can be extracted  from measurements of the decay length and the invariant mass of the products,
even in presence of missing energy.
}

\preprint{MIT-CTP-4142}

\newcommand{\nn}{\nonumber}
\newcommand{\be}{\begin{equation}}
\newcommand{\ee}{\end{equation}}
\newcommand{\bea}{\begin{eqnarray}}
\newcommand{\eea}{\end{eqnarray}}
\newcommand{\gev}{\textrm{ GeV}}
\newcommand{\tev}{\textrm{ TeV}}

\begin{document}
\section{Introduction}

Dark Matter is an outstanding evidence of physics beyond the Standard Model, and may well be within reach at colliders.  The best known parameter about Dark Matter (DM) is its abundance in the Universe. Other than that, limits  on its mass and interactions with nuclei are extracted from negative results of direct and indirect searches. The theoretical bias is toward a particle of mass of hundreds of GeV with weak-strength interactions.
However, even if the DM particle could be produced at colliders, its presence is inferred indirectly, in the form of missing energy.

In this paper, we propose a simple scenario in which DM leaves a clear trace at colliders. 
We extend the Standard Model (SM) to include a new fermion $\Psi$, having both Dirac and Majorana masses. 
The introduction of a Majorana mass 
has the effect of removing the degeneracy of the two Weyl spinors composing $\Psi$.
The Dirac mass is assumed to be of the order of the electroweak scale, whereas
the Majorana contribution is much smaller, and controls the splitting between the two Weyl states.
In this way, $\Psi$ is very close to be a pure Dirac fermion, hence the name  \textit{pseudo-Dirac}  fermion.
This situation is technically natural since a small Majorana mass can be protected by an accidental $U(1)$ symmetry  and 
arise at loop level \cite{natural}. Therefore, with a Dirac mass of the order of few hundreds of GeV, we would expect a splitting in the range of 1-10 GeV.

The light state is stable and constitutes the DM of the Universe, while 
 the slightly heavier state can  decay  to the light one and produce SM particles.
We call this situation \textit{pseudo-Dirac Dark Matter} (pDDM).
In this setup, the decay length of the heavier to the lighter particle turns out to fit within collider detectors, once the relic abundance measurement is imposed. Thus, pDDM leaves a trace in the form of a measurable displaced vertex.  

Pseudo-Dirac dark matter has intermediate features with respect to the limiting Dirac and Majorana cases.
As for Dirac DM, the relic abundance of pDDM is driven by  efficient  $s$-wave coannihilations between the two nearby states.
On the other hand,  pDDM effectively behaves as a Majorana particle in direct detection experiments,
because the  momentum transfer in  DM-nucleus scatterings is much smaller than the 
splitting between the DM and the heavier state.
Nevertheless, there may be a small contribution to the spin-independent cross section with nuclei due to a slightly non-pure Majorana nature of the mass eigenstates.

We have performed a model independent analysis of pDDM, and then specialized it to
the supersymmetric pseudo-Dirac Bino \cite{muless,supersoft}.
Section \ref{sec:structure} introduces the general setup,
 following an effective field theory approach to classify the relevant interactions of the
pseudo-Dirac particle to the SM. 
In Section \ref{sec:DM} we discuss the bounds for DM, and compute the relic abundance and the direct detection cross section.
We then turn to the collider signals in Section \ref{sec:pheno}, where we  explore  the experimental
prospects of determining the overall scale and the mass splitting in the pDDM sector.
Then, in Section \ref{sec:connection} we combine all these elements together and  draw an interesting connection between the abundance of DM in the Universe and the parameters measurable  at colliders.
Conclusions are drawn in Section \ref{sec:conclusions}, where we also provide an outlook on
possible extensions to the present work.

\section{Structure of the Model}
\label{sec:structure}

\subsection{General structure}

We consider a fermion $\Psi$, singlet under the SM gauge group, with the most general Dirac and Majorana masses,  
\be
\mathscr{L}_{\,0}=
\bar\Psi (i\slashed{\partial}-M_D)\Psi - {m_L\over 2}(\bar\Psi^c P_L \Psi+
\textrm{h.c.}) - {m_R\over 2}(\bar\Psi^c P_R \Psi+
\textrm{h.c.})\, ,
\label{lagrangian1}
\ee
where  $P_{R,L}=(1\pm \gamma^5)/2$. The mass eigenstates are a linear combination of $\Psi$ and $\Psi^c$. In this study we are focusing on a pseudo-Dirac situation, where Majorana masses are suppressed with respect to the Dirac mass term. At zeroth order in $\delta\equiv (m_L-m_R)/M_D\ll 1$, the mass eigenstates $\chi_1$ and $\chi_2$ are given by
\bea
\chi_1& \simeq & {i\over \sqrt{2}}(\Psi-\Psi^c)\,, \\
\chi_2& \simeq & {1\over \sqrt{2}}(\Psi+\Psi^c ) \, .
\eea
Note that the fields $\chi_{1,2}$ are self-conjugates at the zeroth order in $\delta$, i.e. 
$\chi_{1,2} = \chi_{1,2}^c+ {\cal O} (\delta)$. In terms of the mass eigenstates, the lagrangian (\ref{lagrangian1}) can be written as
\be
\mathscr{L}_{\,0}=
{1\over 2}(\bar\chi_1 i\slashed{\partial}\chi_1+\bar\chi_2 i\slashed{\partial}\chi_2)
-{1\over 2}m_1\bar\chi_1\chi_1-{1\over 2}m_2\bar\chi_2\chi_2 \, ,
\ee
where the masses  $m_{1,2}=M_D\mp m + {\cal O} (\delta^2)$ ($m\equiv(m_L+m_R)/2$) are just split by the Majorana term. 
Instead of the three mass parameters of the free lagrangian (\ref{lagrangian1}) we will use the  
set of $m_1, \Delta m\equiv m_2-m_1$ and $\delta$. 

To account for the stability of the DM particle, we introduce a new parity such that $\Psi$ is odd and the whole  SM sector is even. In  this case, the pseudo-Dirac fermion can interact with the SM fermions via non-renormalizable interactions such as
\be
\mathscr{L}_{\textrm{int}}=
{1\over \Lambda^2}\bar\Psi\gamma^\mu(c_L P_L+c_R P_R)\Psi\,
\bar f \gamma_\mu (c_L^{(f)}P_L+c_R^{(f)}P_R)f \, ,
\ee
or, in terms of the mass eigenstates  $\chi_1$ and $\chi_2$, 
\bea
\mathscr{L}_{\textrm{int}}&=&{1\over \Lambda^2} 
\left[{i\over 2}{(c_R+c_L)}\bar\chi_1\gamma^\mu\chi_2+
{1\over 4}{(c_R-c_L)}
\left(\bar\chi_1\gamma^\mu\gamma^5\chi_1+\bar\chi_2\gamma^\mu\gamma^5\chi_2\right)
\right] \nn\\
&&\times\left[\bar f\gamma_\mu(c_L^{(f)} P_L+c_R^{(f)}  P_R) f\right] .
\label{interactions}
\eea
The first term is responsible for the decay of $\chi_2$ into $\chi_1$ and SM fermions 
and its collider phenomenology. It is also responsible for an $s$-wave 
contribution to the annihilation cross section. 
Furthermore, in the first line of Eq.~(\ref{interactions}) there should also be vector operators of the type $\bar\chi_1 \gamma_{\mu} \chi_1$ which arise as a consequence of the non-Majorana, $\delta$-suppressed pieces in $\chi_{1,2}$. Those interactions will only play a role in the discussion of direct detection constraints in  Sect.~\ref{sub:detection}.
Other dimension-6 operators of the kind $m_f \bar\Psi\Psi \bar f f/\Lambda^3$ are suppressed with respect to the ones in Eq.~(\ref{interactions}) by $m_f/\Lambda\ll 1$, $\Lambda$
above the weak scale, and hence they
are negligible (the only exception may be for the top quark, but we exclude this possibility, 
as discussed later).
There also exists a dimension-5 operator $\bar\Psi\Psi H^\dagger H /\Lambda$, 
coupling the fermion $\Psi$ to the Higgs doublet $H$~\cite{hk}. 
This operator leads to velocity-suppressed contributions to the annihilation
cross section of the $\chi_i$'s, having small impact on our analysis; 
thus, we will ignore this operator in the following.
Note also that, although we only consider here the possibility of a spin-1/2 new particle,  a similar analysis  applies  to the case of a complex scalar which splits into two quasi-degenerate real scalars.

The  dimensionless coefficients $c_{R,L}, c_{R,L}^{(f)}$ are model dependent. Generically, 
the four-fermion operators in Eq.~(\ref{interactions}) may be the result of integrating out a heavy 
particle of mass $M$ and electroweak couplings,
\bea
{c_{L,R} \, c^{(f)}_{L,R}
\over\Lambda^2} \sim {g'^2\over M^2}\,.
\label{estimate}
\eea
Notice that assuming  $m \ll M_D$ is technically natural once one considers a symmetry which forbids a Majorana mass, such as a $U(1)$ symmetry. A small violation of $U(1)$ symmetry would lead to small Majorana masses.
 
Similar models have been proposed for inelastic Dark Matter \cite{idm, idmcandidates} (iDM), but with an important difference: unlike iDM,
we do not require the mass splitting of the pseudo-Dirac fermion to be 10 -- 100 keV, instead we
consider natural mass splittings of the order of a few GeV.
 Whereas iDM models are designed to explain the  DAMA modulation, pDDM focuses on the collider-cosmology interplay by relating the DM abundance of the Universe with a remarkable collider signature of measurable displaced vertices.

\subsection{A SUSY realization: the pseudo-Dirac Bino}
\label{susycase}
 
 Supersymmetry provides a natural scenario for pseudo-Dirac fermions. Dirac gauginos arise as a consequence of the $U(1)_R \subset SU(2)_R$ symmetry in ${\cal N}$=2 Supersymmetry \cite{randall}. Breaking of the $U(1)_R$ symmetry generates a small Majorana mass along the lines of the previous section, $m \ll M_D$ in Eq.~(\ref{lagrangian1}).
In the MSSM, gauginos are Majorana particles, i.e.  $M_D=0$. However, Dirac gauginos help to solve or alleviate various problems in Supersymmetry, such as the $\mu$ problem \cite{muless}, naturalness \cite{muless, supersoft}, excessive contributions to the muon anomalous magnetic moment and proton decay \cite{muless}, and flavour problems \cite{muless, supersoft, MRSSM}. Moreover, small Majorana masses would naturally arise as a consequence of suppressed $U(1)_R$ breaking terms -- see for example Ref.~\cite{muless} for a discussion.
  
In practice, Dirac gauginos can be thought of as an extension of the MSSM, where each Majorana gaugino marries a new particle in the adjoint representation of the gauge groups (see Ref.~\cite{Belanger} for a comprehensive study of this situation). For example, one would combine each $SU(2)_L\times U(1)_Y$ gaugino ($\tilde W$, $\tilde B$) with a partner ($\tilde W'$, $\tilde B'$) with terms such as
\bea
-\mathscr{L}&\supset&  
M_1^{\textrm{D}} \tilde B\tilde B'
\,.
\label{lagr1}
\eea
This Dirac mass is $U(1)_R$ preserving, whereas  Majorana mass terms such as 
\bea
-\mathscr{L}&\supset& {1\over 2}m_1\tilde B\tilde B+{1\over 2}m'_1\tilde B'\tilde B'
\,,
\eea
would be suppressed by an approximate $U(1)_R$ symmetry. Indeed, in the limit where  $U(1)_R$ and electroweak symmetry is preserved, the 6 Weyl spinors can be paired up in terms of 3 Dirac spinors, e.g.  $(\tilde B, \tilde B')$.  
One particularly interesting situation is Bino Dark Matter. If $M_1^D<M_2^D, \mu$ , the two lightest states
are a linear combination of $\tilde B$ and $\tilde B'$
\be
\tilde\chi_{1,2}^0= \frac{1}{\sqrt{2}} \left( (1\pm \epsilon)\tilde B+ (1 \mp \epsilon)\tilde B' \right)
\,,
\ee
with  $\epsilon=(m_1-m_1')/2 M^D_1$.
The effect of EWSB and Majorana masses is to split the Bino
into two nearly degenerate Majorana particles, with masses $m_{\tilde B}, m_{\tilde B'}$ \cite{hsieh, leptophil1}. 
Note the field $\tilde B'$ does not interact directly
with matter multiplets and therefore, $\tilde \chi_{1,2}^0$  
 interact with matter only through their $\tilde B$ component.

\section{Dark Matter Bounds }
\label{sec:DM}

The  behaviour of Pseudo-Dirac Dark Matter differs greatly from the more standard limiting  cases of
 pure Majorana or  pure Dirac Dark Matter \cite{hk}.
 On one hand, the Dirac nature of pDDM drives the relic abundance computation, where its unsuppressed $s$-wave coannihilations reduce the DM density.
On the other hand, direct detection bounds are driven by the Majorana component of pDDM, and contributions to the spin-independent cross section are only possible if $\delta \neq 0$.

In this section, we will show how the relic abundance measurement would translate into a constraint on the scale of the interactions in Eq.~(\ref{interactions}), whereas direct detection bounds would lead to a constraint on the Majorana masses.

\subsection{Relic Abundance}
\label{sec:relic}

 When there are two almost-degenerate states, coannihilations are important. 
To deal with the coannihilation of two states with different masses ($m_1$ and  $m_1+\Delta m$) and equal number of degrees of freedom  it is convenient to define an 
effective thermally averaged annihilation cross section as \cite{griest}
\be
\langle\sigma_{\textrm{eff}}v\rangle=
{1\over (1+\alpha)^2}
\left[\langle\sigma_{11}v\rangle+
2\alpha\langle\sigma_{12}v\rangle
+\alpha^2\langle\sigma_{22}v\rangle\right]\,,
\ee
where $\alpha=\left(1+{\Delta m/ m_1}\right)^{3/2}e^{-x {\Delta m/  m_1}}$
and $\langle\sigma_{ij}v\rangle$ refers to the thermal average of the  cross sections for the annihilation of the states $i$ and $j$ times their relative velocity.
For the interactions in Eq.~(\ref{interactions}),
the thermally averaged annihilation cross sections into massless fermions $f$ in the non-relativistic limit are
\bea
\langle\sigma_{ii}v\rangle&=&\sum_f\langle \sigma(\chi_{i}\chi_{i}\to f\bar f)v \rangle=
{1\over 2\pi}\sum_f{\left|{c_R-c_L\over 4}\right|^2}
 \left[|c_R^{(f)}|^2+|c_L^{(f)}|^ 2\right]{m_{i}^2\over \Lambda^4}{1\over x_F}\,,\\
\langle\sigma_{12}v \rangle&=&\sum_f\langle \sigma(\chi_1\chi_2\to f\bar f) v \rangle=
{1\over 8\pi}\sum_f{\left|{c_R+c_L\over 2}\right|^2} 
\left[|c_R^{(f)}|^2+|c_L^{(f)}|^2\right] {(m_1+m_2)^2\over \Lambda^4}\,,
\label{sigma12}
\eea
where $i=1,2$ and $m_2=m_1+\Delta m$. 
 The freeze-out temperature $T_F=m_{1}/x_F$ is determined implicitly by 
\be
x_F=25+\log\left[ {d_F\over \sqrt{ g_* x_F}} m_{1}\langle\sigma_{\textrm{eff}}v\rangle\,  \,6.4\times 10^6 \gev\right]\,,
\ee
where $d_F=2$ is the number of degrees of freedom of the $\chi_i$'s and $g_*=96$ are the number of relativistic degrees of 
freedom at $T_F$. Finally the relic abundance is given by
\be
\Omega_{\textrm{DM}} h^2={8.7 \times 10^{-11} \gev^{-2}\over \sqrt{g_*}\int_{x_F}^\infty dx{1\over x^2}\langle\sigma_{\textrm{eff}}v\rangle }\,.
\label{relicabundance}
\ee

Notice that Majorana-type $\chi_1\chi_1$ and $\chi_2\chi_2$ annihilations are velocity-suppressed, while the Dirac-type 
$\chi_1\chi_2$ is not.
To a good approximation  we can neglect the  velocity-suppressed self-annihilations and  restrict ourselves to  the leading contribution from  coannihilations to the effective cross section, which can be written as
\be
\langle\sigma_{\textrm{eff}} v\rangle ={2\alpha\over (1+\alpha)^2} {C^4\over 8\pi} {(2 m_1+\Delta m)^2\over \Lambda^4}
\label{sigma12app}
\,,
\ee
where we have defined the  combination of dimensionless coefficients of the interaction operators in (\ref{interactions}) 
\be
C^4\equiv {1\over 4}\sum_f \left|c_L+c_R\right|^2 (|c^{(f)}_L|^2+|c^{(f)}_R|^2)\,.
\label{Cdef}
\ee
The sum over fermions in $C$ is restricted to those species which are relativistic at $T_F$.
Since we consider $m_1$ of order few hundredths of GeVs,  the third-generation quarks are excluded from the sum.

As we can see, the cross section  $\langle\sigma_{12} v\rangle$, and hence the leading contribution to the relic abundance, only depends on the masses of the two particles $\chi_1, \chi_2$ and an unspecified mass scale $\Lambda/C$, which encodes
the model dependence 
(see Eq.~(\ref{estimate}) for an estimate of the size). 
The plots in Figure \ref{mvsdeltam} show the regions of correct DM density
in the planes $(\Lambda/C, \Delta m)$ and $(\Lambda/C, m_1)$.

\FIGURE[t]{
\centering
\includegraphics[scale=0.7]{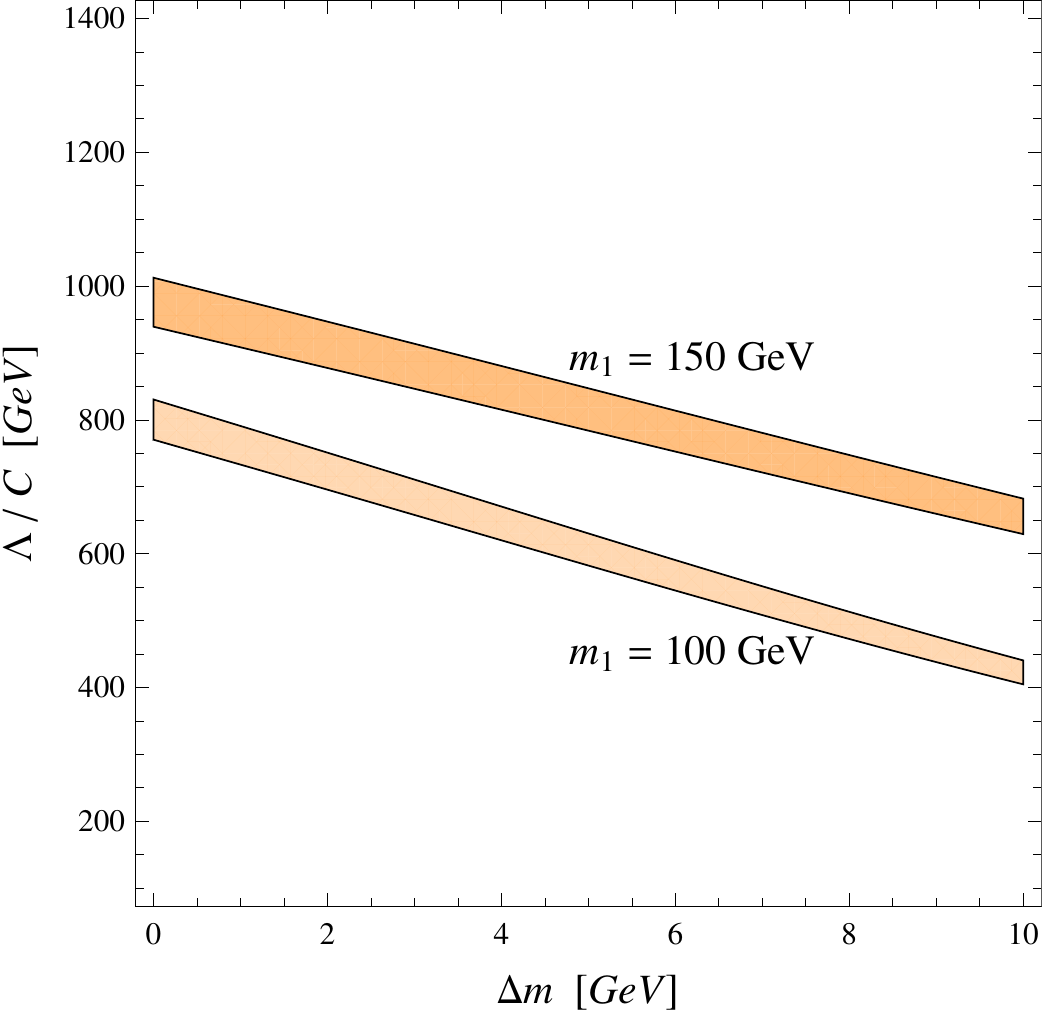}
\includegraphics[scale=0.7]{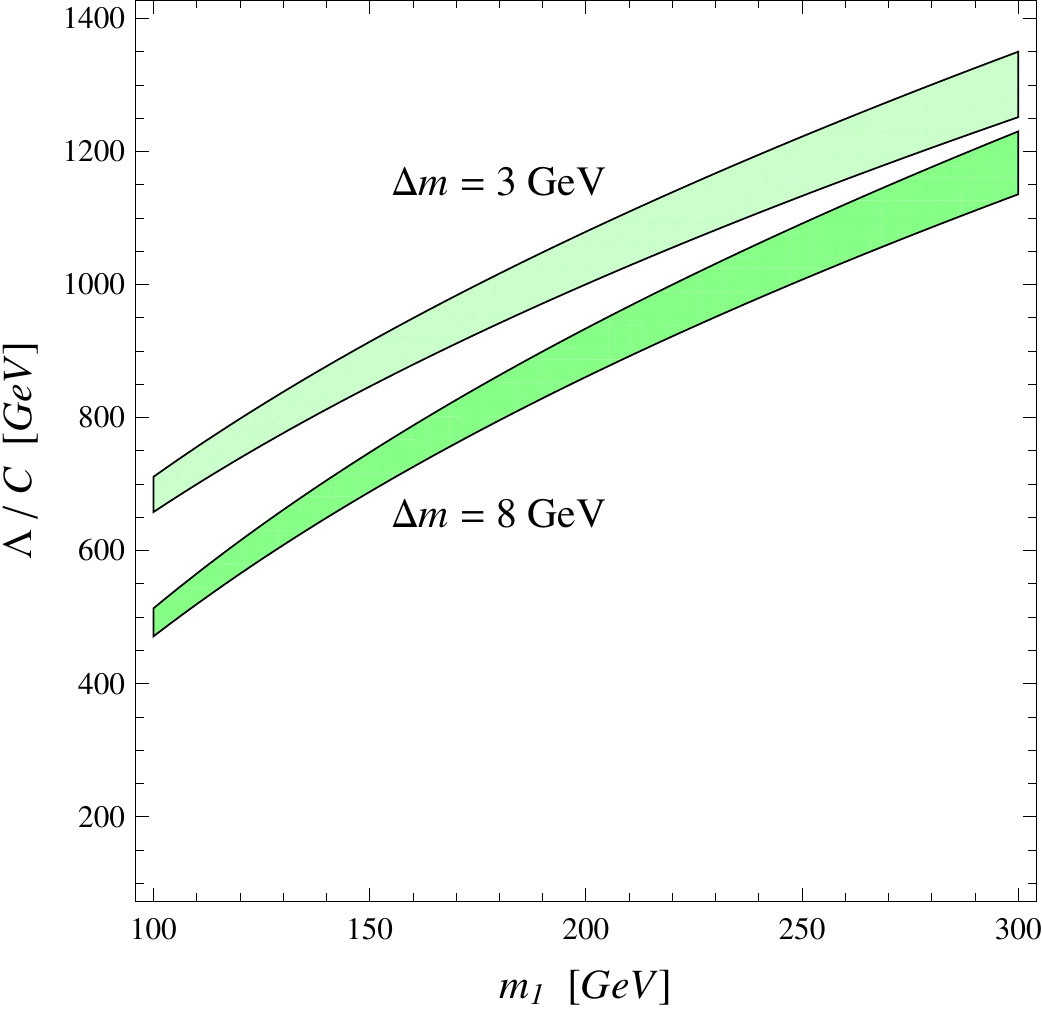}
\caption{Relic abundance 2$\,\sigma$ bands $0.0997<\Omega_{\textrm{DM}} h^2<0.1221$ \cite{wmap7}, for different DM masses $ m_1 =100, 150$ GeV  \textit{(left)} and for different mass splittings $\Delta m=3, 8$ GeV \textit{(right)}. }
\label{mvsdeltam}
}

Inverting Eq.~(\ref{relicabundance}) we obtain a relation between the  parameter $\Lambda/C$ and the relic abundance
\footnote{
We have approximated the integral in (\ref{relicabundance}) using that
\be
\int_{0}^1 dy
{e^{-{x_F\over y}{\Delta m\over m_1}}
\over \left[1+\left(1+{\Delta m\over m_1}\right)^{3/2}e^{- {x_F\over y} {\Delta m\over m_1}}\right]^2}\simeq 
{1\over 4}e^{-x_F {\Delta m\over m_1}}\nn\,,
\ee
valid for $\Delta m/m_1\ll 1$.
},
\be
{\Lambda\over C}\simeq 0.8  \tev \left({\Omega_{\textrm{DM}} h^2\over 0.11}\right)^{1/4}
\left({m_1\over 100 \gev}\right)^{1/2} 
e^{-6{\Delta m\over m_1}}\,.
\label{Lambdaapprox}
\ee
 This approximation agrees  with the numerical findings of Fig.~\ref{mvsdeltam}  to better than 10\%. 

\subsection{Direct Detection}
\label{sub:detection}

Besides the operators written in Eq.~(\ref{interactions}),  there is also a small residual 
vector-vector interaction of the DM to the quarks, due to the 
non-pure Majorana nature of the mass eigenstates.
 The parameter $\delta$ controls these interactions, which may lead to large cross sections with heavy nuclei and 
 thus constraints from direct detection experiments.

In general, the operators relevant to direct detection are the vector-vector and the axial-axial couplings of the DM to the quarks $q$:
\bea
b_q [\bar\chi_1\gamma^\mu\chi_1][\bar q\gamma_\mu q] \,, \qquad 
d_q [\bar\chi_1\gamma^\mu\gamma^5\chi_1][\bar q\gamma_\mu \gamma^5 q]\,,
\eea
where the coefficients  $b_q$, $d_q$ are expressed in terms of the effective lagrangian Eq.~(\ref{interactions}) as 
\bea
b_q= {\delta\over \Lambda^2} B_q^2\,,\qquad \,  d_q= {1 \over \Lambda^2} D_q^2\,,
\eea
with $(B_q^2,D_q^2) \equiv (c_R\pm c_L)(c_R^{(q)} \pm c_L^{(q)})/8$.
Mixed axial-vector and vector-axial interactions give cross sections for direct detection  suppressed by the small DM velocity.

Since the mass splitting between $\chi_1$ and $\chi_2$ is larger than the momentum transfer ($\sim 10-100$ keV) of the DM-nucleus scattering,  direct detection experiments are only sensitive to the lightest state. 
Thus, the usual Majorana DM predictions for spin-dependent cross sections apply here.
The best experimental limits (from Xenon \cite{xenon}) are still above the expected cross sections for Majorana DM \cite{Roszkowski, Essig}  and they are not effectively constraining the parameter space.

On the other hand, the vector-vector term proportional to $\delta$ mediates
coherent spin-independent DM-nuclei scatterings.
The spin-independent total cross section (at zero momentum transfer) of $\chi_1$ on a nucleus with mass $m_N$ of atomic number  
$Z$ and mass number $A$ is
\cite{susydm}
\be
\sigma_0^{\textrm{SI}}={1\over \pi}{m_1^2 m_{N}^2\over (m_1+m_N)^2}\left[ Z b_p+(A-Z) b_n\right]^2
\,,
\ee
where $b_p=2b_u+b_d$ and  $b_n=2b_d+b_u$. 
In the approximation that the DM scatters off neutrons and protons in the same way ($b_p\simeq b_n$), and the DM mass is much 
larger than nucleon mass $m_n$, the cross section per nucleon is given by
\be
\sigma_n^{\textrm{SI}}\simeq \sigma_0^{\textrm{SI}} {m_n^2\over m_N^2}{(m_1+m_N)^2\over (m_1+m_n)^2}{1\over A^2}
\simeq{1\over \pi}{m_n^2}b_n^2\,,
\ee
which can be translated into a constraint on the parameter $\delta$:
\be
\delta\simeq 0.03 \left({\sigma_n^{\textrm{SI}}\over 10^{-43} \textrm{ cm}^2 }\right)^{1/2}
\left({\Lambda/ B_n\over 1 \tev}\right)^2\,.
\label{directdetection}
\ee

The best current direct detection limit on spin-independent DM-nucleon cross section comes from
CDMS-II  \cite{cdms}:
$
\sigma^{\textrm{SI}}_n\lesssim 3\times 10^{-44} \textrm{ cm}^2 \left({m_1/100 \gev}\right)
$, for $m_1 \gtrsim 70$ GeV.
In this case, Eq.~(\ref{directdetection}) can be translated into a bound on $\delta$,
\be
\delta\lesssim 0.02 \left({m_1\over 100 \gev}\right)^{1/2} \left({\Lambda/ B_n\over 1 \tev}\right)^2
\,.
\ee
Therefore, the region of parameter space  dictated by the DM
 relic abundance together with a $\delta$ at the percent level is
 comfortably consistent with direct detection constraints. 
Note  that direct detection  constraints apply to $\delta \propto m_L - m_R$ and not $\Delta m$, and can be relieved by   imposing a parity symmetry relating the $L$ and $R$ sectors.

\section{Collider Phenomenology}
\label{sec:pheno}

In typical scenarios, once  the DM particle  is produced in a collider,
it escapes the detector in the form of missing energy and no other clear feature is available to 
access the production mechanism. Furthermore, the stability of DM usually depends on the conservation of a parity. Therefore, DM candidates are usually produced in pairs, leading to a new challenge for collider physics-- two sources of missing energy which cannot be disentangled.  

In pDDM the situation is dramatically different as there are two quasi-degenerate states, separated in mass by a few GeV. The  heavier  state $\chi_2$ decays into the  lighter  $\chi_1$ with a potentially sizable decay length, allowing for displaced
vertices at the detector. Thus, pDDM leaves a trace at colliders; a visible signature of dark matter production in the form of displaced vertices.

Identifying displaced vertices correctly requires two ingredients: 
1) the decay has to occur  within the tracker (about 1 m), with a minimum length set by the  resolution (about 0.1 mm) \cite{TDRA,TDRC}; 
2) the decay products (typically leptons) should be hard enough to pass  the minimum $p_T$-cuts. For
leptons we conservatively impose a cut of $p_T>4$ GeV.
In this section we will compute the decay length as a function of pDDM parameters and obtain constraints based on detectability of the displaced vertex. Provided the decay products are triggered, a further measurement of dilepton edge provides more information of the pDDM parameters.

\subsection{Decay Length}
\label{sec:length} 

The vector operator in  Eq.~(\ref{interactions}) is responsible of the decay $\chi_2 \to f \bar f \chi_1$. In the limit when $\Delta m \ll m_1$, the width is given by
\be
\Gamma(\chi_2 \to f \bar f \chi_1)\simeq {C'^4 \over 120 \pi^3}{\Delta m^5\over \Lambda^4}\,,
\label{gamma}
\ee 
where $C'$ is defined as $C$ after Eq.~(\ref{sigma12app}) but now the sum runs over SM fermions whose mass is less than $\Delta m/2$. 
For the range of $\Delta m$ under consideration, the $t$-quark is excluded. Decay into
$b$-quarks may be kinematically allowed in a narrow region at large $\Delta m$, but still suppressed respect to decays to lighter particles, leading to a small branching ratio to $b$-quarks. 
Therefore, we neglect the possible
emission of $b$-quarks, which implies $C'=C$.

The coannihilation scattering relevant for the relic abundance ($\chi_1\chi_2\to \bar f f$) and the decay 
length ($\chi_2\to \chi_1\bar f f$) come from the same effective four-fermion interactions
in Eq.~(\ref{interactions}), and the same combination $C/\Lambda$ appears in both expressions, Eqs~(\ref{Lambdaapprox}) and (\ref{gamma}). 
As we will show in Sect.~\ref{sec:connection}, this translates into a relation between the DM abundance in the Universe and a possible measurement of a displaced vertex at colliders.

Notice that the decay length in the $\chi_2$ rest frame,
\be
 \, L_0=\Gamma(\chi_2 \to f \bar f \chi_1)^{-1} \simeq 4.6 \textrm{ cm}\left({\Lambda/C\over
 500 \gev}\right)^4 \left({1 \gev\over \Delta m}\right)^5\,,
\label{decaylength}
\ee
is related to the decay length in the laboratory frame $L_{\textrm{lab}}$ by
\be
L_{\textrm{lab}}=\frac{p_2}{m_2}\,L_0\,.
\ee
$p_2=|\vec{p_2}|$ is the momentum of $\chi_2$, typically of the order of a few times $m_2$ -- see Sec.~\ref{trace}. Therefore, 
a mass splitting of the order of GeV naturally leads to a decay length of the order of a measurable displaced vertex.

The decay length depends on the strength of the coupling of pDDM to the SM fermions, see Eq.(\ref{interactions}), but it is intriguing that  electroweak couplings and masses of the order of a few hundreds of GeV would lead to an observable displaced vertex. The range of observability at LHC and TeVatron is of the order of 100 $\mu$m to 1 m \cite{TDRA,TDRC}.

\subsection{Leaving a Trace at Colliders}
\label{trace}

To describe the pseudo-Dirac phenomenology at colliders we need to specify the production mechanism.
The four-fermion operator in Eq.~(\ref{interactions}) describes the interaction of pDDM with the SM fermions but does not capture interactions involving new heavy particles besides $\chi_{1,2}$. For example, in a supersymmetric scenario where $\chi_{1,2}$ are neutralinos, the main production mechanism is not given by Eq.~(\ref{interactions}) but instead by pair production of squarks with subsequent decays to neutralinos.

With more generality, one could describe the pair production of colored particles ($G$) which would decay into the pDDM via  $G \rightarrow j + \chi_2$. $\chi_2$ decays would  be driven by the interaction in Eq.~(\ref{interactions}),
\be
\chi_2 \rightarrow f \, \bar f \,  \chi_1  \label{3b}\,,
\ee
where $f$ is a lepton or a jet.

In the rest frame of $\chi_2$,  a small $\Delta m$ implies that the $p_T$ distribution of the leptons or jets is small, typically $p_T^{f} < \Delta m$. Objects with very low $p_T$ would not be triggered \cite{TDRA,TDRC}, hence, a sizable boost from the $\chi_2$ reference frame to the laboratory frame is a requirement for detection. Typically, $\chi_2$ would have a $p_T$ of order of the mass of the heavy colored particle \cite{expstudy}. LEP energies are too low to produce such a boost, as pointed out in
Ref.~\cite{threebody}, but at TeVatron or LHC $\chi_2$ would typically carry the $p_T$ of  the heavy parent $G$, and the leptons or jets could have a sizable $p_T$.  

To determine whether such a boost would render the leptons or jets detectable, and with what efficiency, we performed a Monte Carlo simulation with MadGraph/MadEventv4.3 \cite{MG}.  As we mentioned, the dominant decay mechanism in the supersymmetric example would be the production of a squark which decays to the heaviest neutralino, $\tilde{q} \rightarrow \tilde\chi_2^0 + j$, and the subsequent decay of $\tilde \chi_2^0$ into 2 SM fermions and the lightest neutralino.
 
In the rest of this section we are going to consider decays to leptons, $f=e, \mu$, although the discussion can be generalized to any scenario of pDDM where $\chi_2$ has some branching ratio to two leptons. In this case, the final state we are considering is 
\be
\textrm{2 hard jets + 4 leptons + }  \slashed{E}_T \,.
\label{final} 
\ee
This signal contains many leptons, high-$p_T$ jets and missing energy and therefore the background is reducible \cite{leptosusy,matching} and the measurement is not very sensitive to a good determination of the Standard Model background\footnote{In fact, the most sizable background would come from fakes 
\cite{leptosusy,matching}, e.g., jet faking an electron.}. In pDDM, a large missing energy ($\slashed{E}_T \geq 200$ GeV) and two high-$p_T$ jets would be the main handles for triggering. Off-line reconstruction of the decay length would be possible as long as the leptons are good quality leptons, see Fig.~\ref{pTdis} for the efficiency.

\FIGURE[t]{
\centering
\includegraphics[width=8.5cm, height=6.5cm]{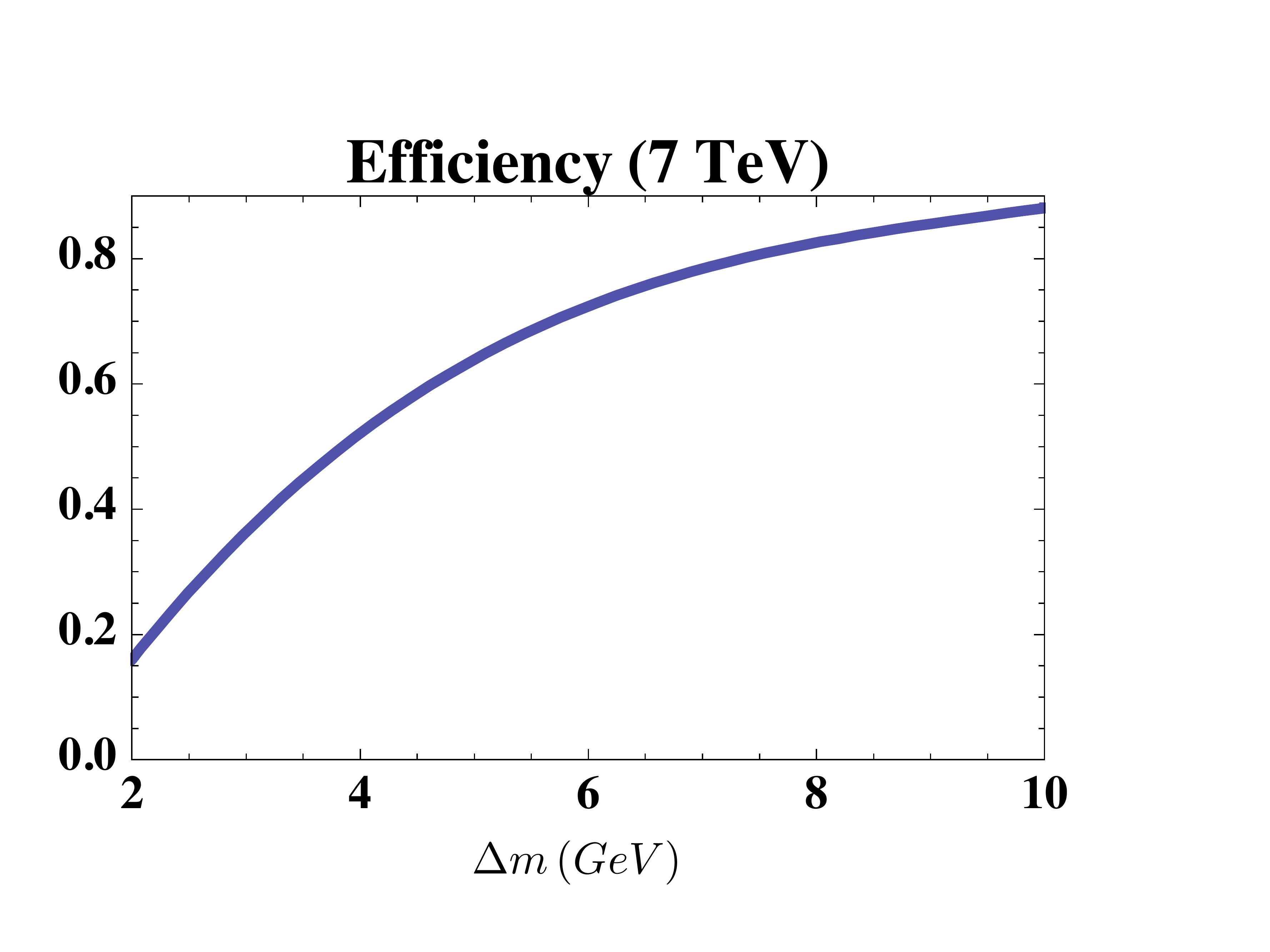}
\hspace{-1cm}
\includegraphics[width=6cm, height=7.5cm, angle=90]{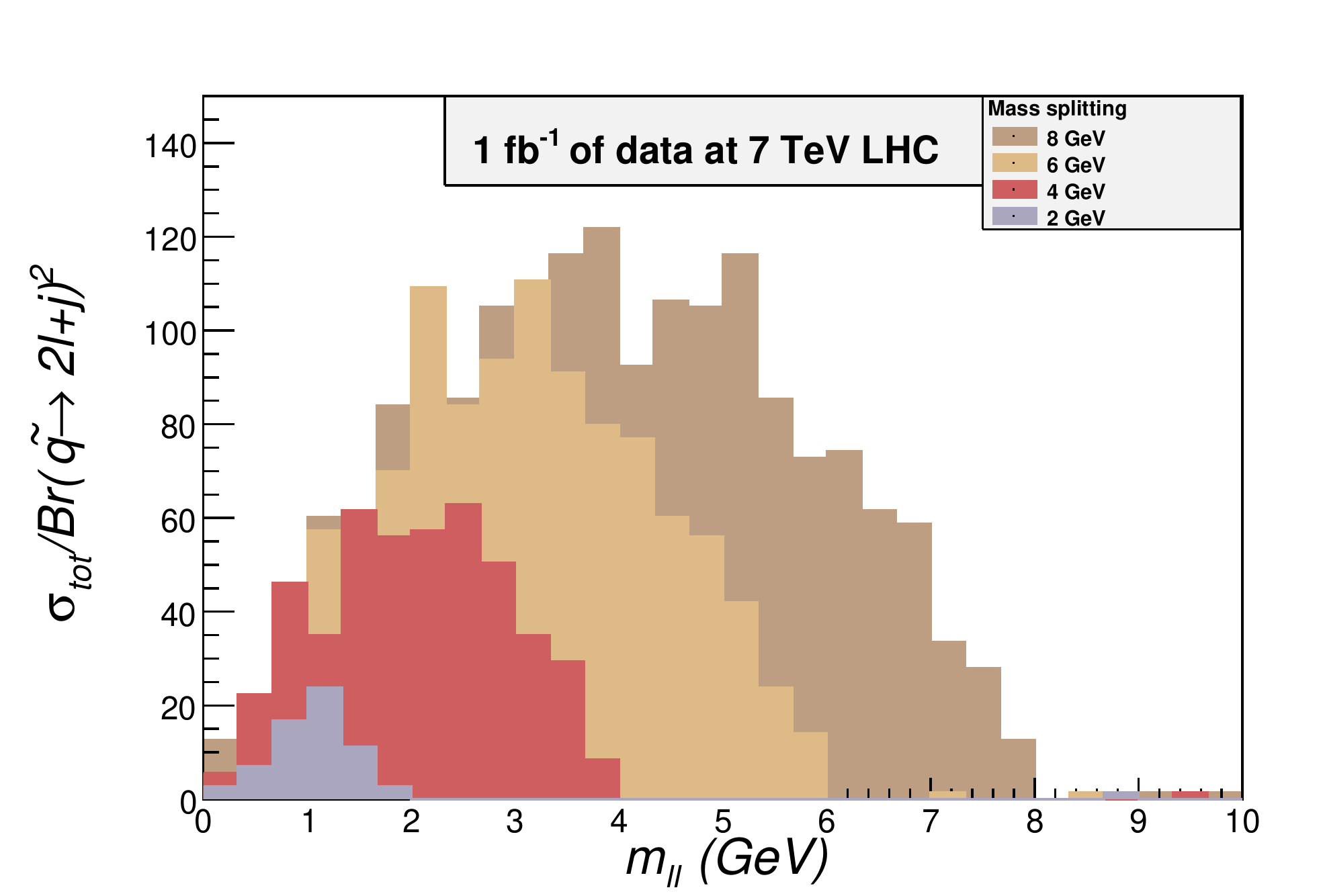}
\caption{{\it Left:} Efficiency of minimum $p_T$ cut on leptons as a function of $\Delta m$. Both plots are generated with a simulation of LHC 7 TeV run. {\it Right:} Dilepton invariant mass distribution for different choices of $\Delta m$ (LHC 7 TeV run).}
\label{pTdis}
}

In pDDM, two measurements, the dilepton edge and decay length, suffice to determine the overall DM scale and the splittings.
 The endpoint in the dilepton invariant mass distribution provides a measurement of $\Delta m$. Indeed, if the three-body decay proceeds through the exchange of a heavy particle, such as a slepton,  the edge depends on the mass difference between the two neutral states
\be
m_{\ell \ell}^{edge} = \Delta m =  m_{\chi_2^0}-m_{\chi_1^0},
\ee
as discussed in Ref.~\cite{expstudy} and in Sec.~``Measurements from Supersymmetric Events" of
Ref.~\cite{TDRA}. 

To measure this edge we asked for 4 leptons with $p_T>$ 4 GeV and paired them asking for the closest 2 opposite-sign leptons \cite{leptosusy}. After this selection, the combinatorial background is very small, and the dilepton invariant mass has a clear edge at the position of $\Delta m$, see right side of Fig.\ref{pTdis}\footnote{We also asked for leptons with $|\eta|< 3.5$ and minimum separation between leptons of $\Delta R_{\ell \ell}> 0.7$.}.  The lepton momentum can be determined with a precision of a few percent for $p_T \geq 4$ GeV\cite{TDRA,TDRC}, hence a determination of $\Delta m$ is possible within the range of a few percent. 
 In the example shown in Fig.\ref{pTdis} we chose 500 GeV squarks decaying into $\chi_2^0$ of mass $100+\Delta m$ GeV, and then we varied $\Delta m$. Note, though, that there are no LEP bounds for neutralino masses when $\Delta m <$ 4 GeV, and smaller neutralino masses could be considered.  The $p_T$ distribution depends on the energy of the collider, and we simulated the events assuming a 7 TeV running for the LHC. We decayed $\chi_2$ with BRIDGE \cite{bridge}. In the left side of Fig.~\ref{pTdis}, the efficiency of a $p_T$ cut of 4 GeV for at least two leptons is shown in Fig.(\ref{pTdis}). With a splitting between the two Majorana states $\chi_1$ and $\chi_2$ of 2 GeV, the efficiency is 20\%, whereas for 10 GeV the efficiency is close to 100\%. The efficiency depends on the energy of the collider, and Fig.\ref{pTdis} corresponds to LHC at 7 TeV. 
Although TeVatron has no kinematic access to 500 GeV squarks, we could consider lighter squarks. The efficiency of the $p_T >$ 4 GeV cut would be lower, as $\tilde\chi_2^0$ would be less boosted, $p_{T,\tilde\chi_2^0}$ $\sim m_{\tilde q}/2$.
 
One could obtain more information on the spectrum by pairing the two leptons with the nearest jet. See \cite{allanach} for a discussion on the $\ell \ell j$ kinematic edges\footnote{Note, though, that their analysis applies to a sequential decay, where the slepton is lighter than $\chi_2^0$.}.

The other essential ingredient to pDDM at colliders is the measurement of a displaced vertex. As we discussed in Sec.~\ref{sec:length}, the proper length $L_0$ and the length measured in the laboratory differ by a factor $p_2/m_2$, but one can measure $p_2$ as follows. Bounds on squarks -- or any coloured particle -- at TeVatron indicate that $m_{\tilde q} \gtrsim 400$ GeV \cite{TeVonbounds, Abazov:2009rj}. Such heavy particles would be produced with very little boost at a 7 TeV collider, resulting in a back to back jet and $\chi_2$, i.e. $p_2 \sim p_j$. We have verified this statement with the Monte Carlo simulation in our example of 500 GeV squarks. To summarize, a measurement of the jet momentum would provide a good estimate for the 3-momentum of $\chi_2$ and the coloured particle mass. That estimate would translate into a measurement of the pDDM width, which can be directly related to the  DM density
in the Universe, as we now turn to show.

\section{The Dark Matter -- Collider Connection}
\label{sec:connection}

\begin{figure}[t]
\centering
\includegraphics[scale=0.7]{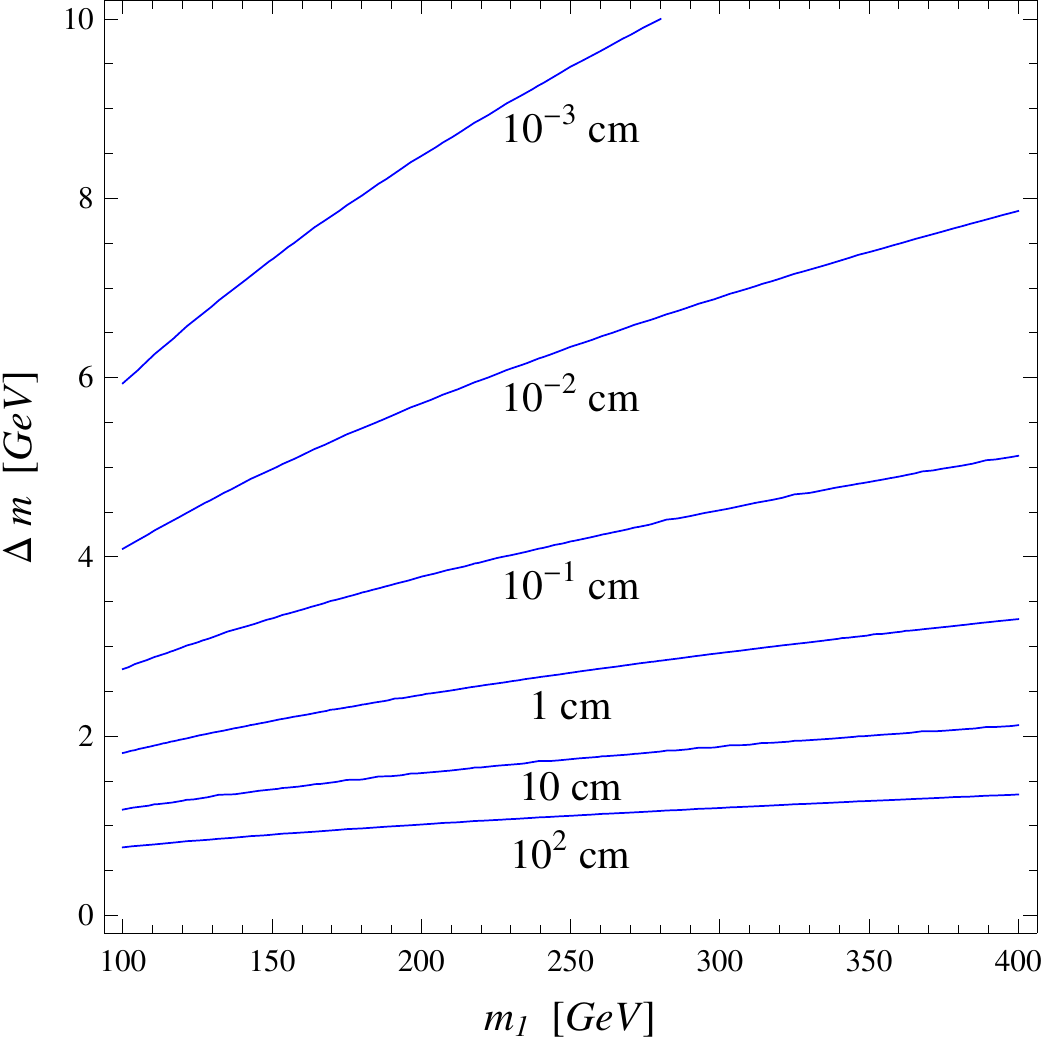}
\caption{Proper decay lengths for $\chi_2\to f\bar f\chi_1$ in the plane of the mass splitting $\Delta m$ between  $\chi_1$ and $\chi_2$ and the DM mass $m_1$. 
The   relic abundance of $\chi_1$ has been fixed to $\Omega_{\textrm{DM}}h^2=0.11$.}
\label{lengths}
\end{figure}

Having discussed in the previous sections the DM bounds and the 
collider phenomenology of pDDM, we are now in a position
to connect  these very different pieces of information. 
We can eliminate $\Lambda/C$ in Eq.~(\ref{decaylength}) using Eq.~(\ref{Lambdaapprox})
and obtain an expression for the decay length of $\chi_2\to f\bar f \chi_1$, which includes the relic abundance:
\be
L_0\simeq 30 \textrm{ cm} \left({\Omega_{\textrm{DM}} h^2\over 0.11}\right)
\left({m_1\over 100 \gev}\right)^{2}\left({1 \gev \over \Delta m}\right)^{5} 
{e^{-{24}{\Delta m\over m_1}}}\,.
\label{Lfinal}
\ee
This relation is the main result of the paper. It provides an intriguing connection
between cosmological and collider measurements, which is independent of the details of
the coefficients of the effective lagrangian, and only contains readily measurable quantities.
This  remarkable property of pDDM is easy to understand:  the processes leading to   the relic abundance ($\chi_1\chi_2\to \bar f f$) and  the decay 
length ($\chi_2\to \chi_1\bar f f$) come from the same term in the effective lagrangian
in Eq.~(\ref{interactions}). 
Still, it is an outstanding coincidence that the constraints from DM abundance  and from  having a mass splitting which is loop-suppressed  with respect to the overall DM mass scale, point to the region of parameter space
 corresponding to  visible displaced vertices $10^{-2}  \textrm{ cm}\lesssim L_0\lesssim 10^2 \textrm{ cm}$, see 
Fig.~\ref{lengths}.

It is worth noticing that the simple relation between collider and cosmological quantities also allows to make predictions
which may rule out the model. In fact, suppose that from measurements
of the dilepton invariant mass distribution we extract $\Delta m$. 
The decay length is also easily measured. The relation (\ref{Lfinal}) then makes a prediction
for the dark matter mass $m_1$, which can be tested against other independent measurements
from direct or indirect searches.

As mentioned in Sect.~\ref{susycase}, a concrete realization of the pDDM scenario is
provided by the supersymmetric Bino with a singlet partner. 
Recall that a Majorana pure Bino, as in the MSSM, is typically overproduced for the
range of masses above the LEP limits, due to its small annihilation cross section
(see e.g. Ref.~\cite{welltempered}). Instead, a pseudo-Dirac Bino avoids this problem
because it coannihilates efficiently with the nearby state. 

The translation from our effective lagrangian to the Bino case is straightforward.
From the Bino-fermion-sfermion interactions of the supserymmetric lagrangian,
one can integrate out the heavy sfermion of mass $m_{\tilde f}$ and obtain an
effective lagrangian of the kind in Eq.~(\ref{interactions}). Then it is possible to read
off the coefficients and, for the relevant combination of parameters as in $C$ of  Eq.~(\ref{Cdef}), 
we get
\be
\left({C\over \Lambda}\right)^4=\sum_f {1\over 4}\left({g'\over m_{\tilde f}}\right)^4  Y_f^4\,,
\label{Cbinocase}
\ee
where $g'$ is the U(1)$_Y$ gauge coupling and $Y_f$ is the hypercharge of the fermion $f$.
If the neutralinos $\tilde \chi_{1,2}^0$ are pure combinations of $\tilde B$ and $\tilde B'$, and the decay occurs 
mostly through the exchange of a right-handed slepton of mass $m_{\tilde\ell_R}$, the decay length is obtained from
Eqs.~(\ref{decaylength}) and (\ref{Cbinocase}):
\be
L_0\simeq 1.8 \textrm{ cm}  \left({m_{\tilde \ell_R}\over
 100 \gev}\right)^4 \left({1 \gev\over \Delta m}\right)^5 \ ,
\label{length} 
\ee
valid up to order ${\cal O}(\Delta m/ m_{\tilde\chi_{1,2}})^2$.  
Instead, for DM annihilations only into right-handed leptons,  the analytical approximation in 
Eq.~(\ref{Lambdaapprox}) translates into
\be
{m_{\tilde \ell_R}}\simeq 202 \gev \left({\Omega_{\textrm{DM}} h^2\over 0.11}\right)^{1/4}
\left({m_1\over 100 \gev}\right)^{1/2} 
e^{-6{\Delta m\over m_1}}\,,
\ee
which can be regarded as a  prediction for the slepton mass, once the bottom of the supersymmetric spectrum is known.

\section{Conclusions and Outlook}
\label{sec:conclusions}

In this paper, we presented a scenario called pseudo-Dirac Dark Matter (pDDM) which possesses a virtue uncommon to DM theories:
observable collider signals in the form of displaced vertices.

In pDDM, the DM particle is accompanied by a slightly heavier state.
This may arise as a consequence of an approximate $U(1)$ symmetry. 
The small mass splitting is responsible for the displaced vertices at colliders, characterized by a  very suggestive form of the
decay length:
\be
L_0\simeq 30 \textrm{ cm} \left({\Omega_{\textrm{DM}} h^2\over 0.11}\right)
\left({m_1\over 100 \gev}\right)^{2}\left({1 \gev \over \Delta m}\right)^{5} 
{e^{-{24}{\Delta m\over m_1}}}\,.
\ee
We have described how one could determine $L_0$ using the combination of  displaced vertex and jet momentum measurements. Moreover, one can obtain direct information
on the splitting $\Delta m$ by measuring a di-lepton edge from the decay products. Those measurements combined would lead to a prediction of the DM mass, which can be tested against other independent
measurements from direct or indirect searches.
Thus, pDDM is predictive and easily testable.

Besides the unusual collider signatures, pDDM presents other interesting features: $s$-wave coannihilations between the two nearly-degenerate states drive the DM relic density, whereas  potentially dangerous vector interactions with the quarks in the nuclei are absent or harmless. 

It is remarkable that  a $\Delta m/m_1$ of the order of a loop suppression and the observed relic abundance conspire to predict a decay length
visible at colliders.  The measurement of the decay length  provides a new insight into DM, beyond the usual missing energy distribution information. 

There are several interesting directions to extend the present work, which we leave
for future investigations: the inclusion of the effects of the dimension-5 operator, as mentioned
in the Introduction; the study of the complex scalar case; a more detailed assessment of the 
LHC reach, beyond the parton level analysis.

\acknowledgments
ADS is indebted to Francesco D'Eramo for many useful discussions. VS thanks Wendy Taylor for conversations about ATLAS and TeVatron vertexing capabilities, and Isabel Trigger and Peter Krieger for discussions on neutralino LEP bounds, and Pierre Savard and William Trischuk for clarifying triggering issues.  
The work of ADS was supported 
in part by the INFN ``Bruno Rossi'' Fellowship, and in part 
by the U.S.  Department of Energy (DoE) under contract No. 
DE-FG02-05ER41360.  
The work of VS and HS is supported partly by NSERC funding.



\end{document}